\documentstyle[11pt,paspconf]{article}
\input epsf.tex

\def\lsim{\lower0.6ex\vbox{\hbox{$ \buildrel{\textstyle <}\over{\sim}\ $}}}
\def\rsim{\lower0.6ex\vbox{\hbox{$ \buildrel{\textstyle >}\over{\sim}\ $}}}

\def\edcomment#1{\iffalse\marginpar{\raggedright\sl#1\/}\else\relax\fi}
\reversemarginpar

\begin{document}
\title{Testing Big Bang Nucleosynthesis}
\author{Gary Steigman}
\affil{Departments of Physics and Astronomy, The Ohio State University,
174 West 18th Avenue, Columbus, OH 43210, USA}

\begin{abstract}
Big Bang Nucleosynthesis (BBN), along with the cosmic background 
radiation and the Hubble expansion, is one of the pillars of
the standard, hot, big bang cosmology since the primordial
synthesis of the light nuclides (D, $^3$He, $^4$He, $^7$Li)
must have occurred during the early evolution of a universe
described by this model.  The overall consistency between the
predicted and observed abundances of the light nuclides, each of
which spans a range of some nine orders of magnitude, provides
impressive support for the standard models of cosmology and
particle physics.  Here, the results of recent, statistically
consistent tests of BBN are described.  This new confrontation
between theory and data challenges the standard model.  The crises confronting BBN are identified and several possible
resolutions are outlined.
\end{abstract}

\section{Introduction}
The discovery of the Cosmic Background Radiation (CBR) by Penzias \& Wilson
(1965)  transformed forever the study of Cosmology from an exercise in
philosophy to the pursuit of science.  The presence of the CBR in an expanding
Universe favors the hot big bang cosmology.  A Universe described by this
model was very hot and very dense during early epochs in its evolution.  As a
consequence, it is a prediction of this ``standard" cosmological model that,
briefly, the early Universe was a primordial nuclear reactor in which the
light nuclides D, $^3$He, $^4$He and $^7$Li were synthesized in astrophysically
interesting abundances (for details and references see, e.g., Boesgaard \&
Steigman 1985;  Walker et al. 1991).  Thus, along with the CBR and the ``Hubble" expansion,  Big Bang Nucleosynthesis (BBN)
provides one of the three
pillars supporting the standard model of cosmology.  The standard hot big
bang model is, in principle, falsifiable.  In contrast to cosmology
as theology, this empirical model is not a matter of faith but, rather,
demands our eternal vigilance and critical scrutiny.    The success of the model is gauged
by the degree to which the BBN predictions are consistent with the primordial
abundances of the light nuclides inferred from observational data.  Over the
years BBN has emerged unscathed from the confrontation between theory and
observations, providing strong support for the standard, hot big bang
cosmological model (e.g., Yang et al. 1984;  Boesgaard \& Steigman 1985;
Walker et al. 1991).  This success has, however, not spawned complacency, and
the testing continues.  In recent years, as the astronomical data has become
more precise, hints of a possible crisis have emerged (Copi, Schramm \&
Turner 1995; Olive \& Steigman 1995;  Hata et al. 1995).  It is my goal here
to describe the impressive success of BBN and to map out the paths leading to
the current challenges to the standard model.  To better appreciate these
challenges and the opportunities they present, an historical analogy may be
instructive.

\subsection{Three Crises For The 19th Century Standard Model}
The gravitational theory described by Newton (1686) was
outstandingly successful in explaining the motion of the moon and planets.
Perhaps one of the most thoroughly tested physical theories in history,
Newtonian gravity had become the standard model of 19th Century Physics.
Soon, however, some challenges to the standard model emerged.  The nature of
these challenges and their different resolutions provide some interesting
lessons for the emerging crisis in BBN.
\vskip .2in
\noindent (i)~~Perturbations to the Orbit of Uranus

Deviations in the orbit of Uranus from the predictions of Newtonian gravity
(the standard model) led Adams and LeVerrier to predict the existence and
location of Neptune.  The standard model was used to discover something new,
verifying the accuracy of the data (the orbit of Uranus) and providing
spectacular support for Newtonian gravity.
\vskip .2in

\noindent (ii)~~Perturbations to the Orbit of Neptune

Observations of the newly discovered Neptune suggested that its orbit, too,
was being perturbed away from the standard model predictions.  So began the
long search which culminated in the discovery of Pluto.
Pluto, however, is not responsible for measurable perturbations to the orbit
of Neptune - it is too small.  Rather, here we have a case of insufficiently
accurate data.  The discovery of Pluto was serendipitous;  more accurate
observations of Neptune's orbit are entirely consistent with the predictions
of Newtonian gravity.

\vskip .2in

\noindent (iii)  Precession of the Perihelia of Mercury

By the mid-19th century LeVerrier had noted a discrepancy between the
predicted and observed precession of the perihelia of Mercury.  LeVerrier
and others proposed one or more planets (Vulcan) between Mercury and the Sun
to resolve this crisis.  None were found.  Alternately, it was proposed by
Newcomb and others that the perturbing mass might be in a ring of dust or
asteroids.  This, however, would have perturbed the orbits of Mercury and
Venus in conflict with observational data.  A more radical solution,
modifying the inverse square law, was proposed by Newcomb (1895).  This,
however, is in conflict with the accurately observed lunar orbit.

As is so well known, the resolution of this crisis confronting the 19th
century standard model was new physics!  Einstein's General Theory of
Relativity (1916) predicts a precession in beautiful agreement with that
observed.
\vskip .2in
\noindent
Three crises, three different resolutions: the standard model preserved and
a new discovery (Neptune);  the standard model preserved and insufficently
accurate data (Neptune/Pluto);  the standard model replaced (perihelia of
Mercury).

\section{Consistency of the Standard Model}

\subsection{Predictions}
Now let us turn to the standard model of cosmology and the predictions of
primordial nucleosynthesis.  Employing measured weak interaction rates and
nuclear reaction cross sections the primordial abundances of the light
nuclides are predicted by BBN as a function of only one adjustable parameter,
$\eta$, the universal ratio of nucleons (baryons) to photons
$(\eta = N_B/N_\gamma; \eta_{10} = 10^{10}\eta)$.  The predicted abundances
of
$^4$He ($Y$ is the $^4$He mass fraction) D and $^7$Li ($y_2 = N_D/N_H,  y_7 =
N_{Li}/N_H)$ are shown for $1 \le \eta_{10} \le 10$ in Figure 1 from Hata et
al. (1995).  For clarity of presentation the predicted abundance of $^3$He,
very similar to that of D, is not shown.

The predicted abundances depend on the universal expansion rate, $t^{-1}$,
during the epoch of BBN ($\sim3 MeV \rsim T_{BBN} \rsim 30 keV; 0.1 \lsim
t_{BBN} \lsim 10^3 sec$).  For the early Universe $t^{-1} \propto
\rho_{TOT}^{1/2}$, where $\rho_{TOT}$ is the total mass-energy density.   For
the ``standard" model (SBBN), $\rho_{TOT}$ is dominated by photons,
electron-positron pairs and three flavors of light, left-handed neutrinos
$(\nu_e,\nu_\mu, \nu_\tau)$.

$$
\rho_{TOT}^{SBBN} = \rho_\gamma + \rho_e + 3\rho_\nu^0.\eqno(1)$$

In (1), $\rho_\nu^0$ is the contribution from one flavor of light $(m_\nu
<<T_{BBN})$ neutrinos.  To account for a possibly massive $\tau$-neutrino
and/or for other, new particles beyond the standard model, it is convenient
to modify eq. (1) by introducing
$N_\nu$, the ``effective" number of equivalent light neutrinos (Steigman,
Schramm \& Gunn 1977).

$$
\rho_{TOT}^{BBN} = \rho_\gamma + \rho_e + N_\nu\rho_\nu^0.\eqno(2)$$
For SBBN, $N_\nu = 3$;  for $N_\nu \neq 3$ the universal expansion rate at
BBN is
modified.  For $N_\nu \ge 3$, the universe expands more rapidly leaving less
time for the conversion of neutrons to protons.  Since most neutrons are
incorporated in $^4$He,  $Y_{BBN}$ increases with $N_\nu$ (and, vice-versa).
Therefore, it is convenient to use
$N_\nu$ as a second parameter to explore deviations from SBBN and extensions
of the standard model of particle physics (Steigman, Schramm \& Gunn 1977).
The results in Figure 1 are for SBBN $(N_\nu = 3)$.

For $1 \le \eta_{10} \le 10$, the predicted abundances of the light nuclides
span a range of some 9 orders of magnitude from $\sim 10^{-10} -10^{-9}$
for
Li/H, to
$\sim 10^{-5}  -10^{-4}$ for D/H and $^3$He/H, to $\sim $ $0.1$
for $^4$He/H.

\subsection{Observations}

Primordial abundances are, of course, not observed.  Rather, they are
inferred from astronomical data.  Some, such as D and $^3$He, have been
mainly observed ``here and now" (in the solar system and the interstellar
medium (ISM) of our own Galaxy).  For these nuclides it is necessary to
extrapolate from here and now to ``there and then" to derive their universal
primordial abundances.  $^4$He and $^7$Li are observed (in addition to here
and now) in regions where much less chemical processing has occurred (low
metallicity, extragalactic HII regions for $^4$He;  very metal-poor halo
stars for
$^7$Li).  For these nuclides the extrapolations to primordial abundances are
smaller.

In addition to observational uncertainties and those associated with the
extrapolations to primordial abundances, systematic effects in deriving
abundances from data may contribute to the overall uncertainties.  The bad
news is that such systematic uncertainties are difficult to constrain.  The
good news is that the sources of possible systematic errors are different for
the different nuclides.

\subsection{Testing SBBN}

The relatively strong and monotonic $y_2$ vs. $\eta$ relation visible in
Figure 1 points to
D (and, to a lesser extent, $^3$He) as an ideal baryometer.  If the
primordial abundance of D were known, for example, to $\sim 40\%$, the
universal density of baryons would be known to $\sim 25\%$.  The large
extrapolation from here and now to there and then has inhibited the
implementation of this approach.  Rather, to avoid this large extrapolation,
a more conservative approach has been adopted.  Since any D incorporated in
a star is burned (to $^3$He) and there are no significant astrophysical
sources of post-BBN D, the abundance of D observed anywhere at anytime
provides a lower bound to its primordial abundance (e.g., $y_{2P} \ge y_{2\odot}$,
$y_{2P} \ge y_{2ISM}$).   From Figure 1 it is clear that a lower bound to $y_{2P}$
leads to an upper bound to $\eta$.

\begin{figure}
\protect\centerline{
\epsfxsize=3.5in\epsffile[ 30 80 580 580 ]{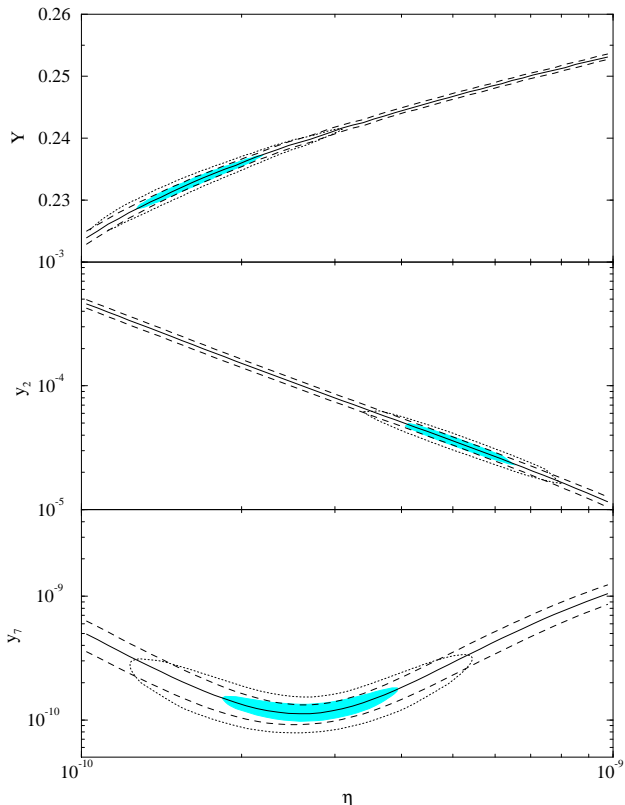}
}
        \caption{The
SBBN predicted abundances (solid lines) of $^4$He (Y is the $^4$He mass fraction), D ($y_2$ = D/H), and $^7$Li ($y_7$ = Li/H) as a function of the nucleon-to-photon ratio $\eta$.  The dashed lines are the $1\sigma$ theoretical uncertainties from  Monte Carlos.  The shaded (dashed) contours are the regions constrained by the observation at the 68\% (95\%) CL.}
       
\end{figure}

It is difficult to avoid the uncertainties of chemical evolution models in
using observations of D to infer an upper bound to $y_{2P}$.  However, Yang et
al. (1984) noted that since
D is burned to $^3$He and some $^3$He survives stellar processing, the
primordial abundances of D + $^3$He are strongly correlated with the
evolved abundances of D +
$^3$He.  Burying the stellar and evolution model uncertainties in one
parameter, $g_3$, the $^3$He survival fraction, Yang et al. (1984;  also,
Walker et al. 1991) used solar system data to place an upper bound on
primordial D (and/or on D + $^3$He).  An upper bound on $y_{2P}$ provides a
lower bound on $\eta$ (see Fig. 1).

Due to the ``valley" shape in the BBN prediction of Li vs. $\eta$ (see Fig.
1), an upper bound to $y_{7P}$ will provide both lower and upper bounds to
$\eta$.  The lithium abundance also offers a key test of the standard model
since its primordial value must not lie below the minimum predicted $(y_{7
BBN} \rsim 1 \times 10^{-10})$.

One test of the consistency of SBBN is to use D, $^3$He and $^7$Li to
infer lower and upper bounds to $\eta$ ($\eta_{MIN}$, $\eta_{MAX}$) and to check
that $\eta_{MIN} < \eta_{MAX}$.  If SBBN passes this test, the ``$^4$He
test" may
be applied.  The predicted $^4$He mass fraction, $Y_{BBN}$, is a very weak
function of $\eta$, increasing from $Y_{BBN} = 0.22$ at $\eta_{10}=1$ to
$Y_{BBN} = 0.25$ at $\eta_{10} = 10$.  Thus, it is key to the success of SBBN
$(N_\nu = 3)$ that for $\eta_{MIN} < \eta < \eta_{MAX}$, $Y_P$ (the inferred
primordial abundance) is consistent with $Y_{BBN}(\eta)$ (the predicted
abundance).

\subsection{Consistency}

Yang et al. (1984) were among the first to carry out a detailed analysis of the
observational data and to implement the tests described above.  From D,
$^3$He and $^7$Li (with $g_3 \ge 1/4$, see Dearborn, Schramm \& Steigman
1986) they found consistency:  $3 \lsim \eta_{10} \lsim 7$, leading to a
predicted
range for $^4$He: $0.24 \lsim Y_{BBN} \lsim 0.26$.  Comparing with the rather sparse
data available, they derived $0.23 \lsim Y_P \lsim 0.25$ and concluded that SBBN
passed the $^4$He test.  They did note that SBBN is, in principle,
falsifiable and
pointed out that if future comparisons should increase $\eta_{MIN}$ and/or
decrease $Y_P$, consistency would require $N_\nu < 3$, modifying the standard
model.

By 1991 uncertainties in the neutron lifetime (as well as its central value)
had been reduced considerably permitting a very accurate prediction of
$Y_{BBN}$ vs. $\eta$ (at the $2\sigma$ level, $Y_{BBN}$ is known to $\lsim \pm
0.001$;  see Thomas et al. 1995).  At the same time there was extensive new
data on lithium (in halo stars) and helium-4 (in extragalactic HII regions).
Applying the above tests, Walker et al. (1991) found $2.8 \le
\eta_{10} \le 4.0$ and $0.236 \le Y_{BBN} \le 0.243$.  From the HII region
data, Walker et al. (1991) derived $Y_P = 0.23 \pm 0.01$ and concluded that
SBBN passed the $^4$He test.  However, they did emphasize, ``that if our
lower bound on $\eta$ were increased from $\eta_{10} = 2.8$ to $\eta_{10} =
4.0$, the window on $N_\nu$ would be closed (for
$Y_{P} \lsim 0.240$)."

\subsection{Crisis?}

Recent applications of the two consistency tests ($\eta_{MIN} < \eta_{MAX}$?
$Y_P = Y_{BBN}$?) have provided hints of a possible crisis (Copi, Schramm \&
Turner 1995;  Olive
\& Steigman 1995;  Hata et al. 1995).  The two ``weak links" are the lower
bound on
$\eta$ inferred from D and $^3$He observations and the upper bound on $Y_P$
derived from the extragalactic HII region data.

It has long been known that the D + $^3$He analysis of Yang et al. (1984) and
Walker et al. (1991) is likely overly conservative.  In both analyses the
synthesis of new $^3$He in low mass stars (Iben 1967;  Rood 1972;  Iben \&
Truran 1978) was neglected.  But,  Rood, Steigman \& Tinsley (1976) had
demonstrated that such production might dominate the primordial (D + $^3$He)
contribution.  Even neglecting this contribution, Steigman \& Tosi (1992) had
followed the evolution of $^3$He in a variety of chemical evolution models
and found more $^3$He survival $(g_3 \rsim 1/2$ rather than $g_3
\rsim 1/4)$ leading to a higher lower bound to $\eta$.  More recently,
Steigman \& Tosi (1995) revisited the ``generic" evolution of D and $^3$He
and, using updated solar system data (Geiss 1993) inferred (for $g_3 \ge 1/4$)
$\eta_{10} \ge 3.1$.  In a more sophisticated implementation of the
``generic" approach, Hata et al. (1996) found (for $g_3 \ge 1/4$)
$\eta_{10} \ge 3.5$.

Although it is still true that $\eta_{MIN} < \eta_{MAX}$, the increasing
lower bound to
$\eta$ increases the lower bound to $Y_{BBN}$.  For $\eta_{10} \ge 3.1$ (3.5),
$Y_{BBN} \ge 0.241 (0.242)$.  An accurate determination of $Y_P$ from
observations of $^4$He in low metallicity extragalactic HII regions is
required for
the $^4$He test.  From their analysis of this data, Olive \& Steigman (1995)
derive
$Y_P = 0.232 \pm 0.003$ where $0.003$ is the
$1 \sigma$ statistical uncertainty.  Thus, at $2 \sigma$, $Y_P^{MAX} <
Y_{BBN}^{MIN}$, failing the $^4$He test.  It should be noted that
$\eta_{MIN}$ and $Y_{BBN}^{MIN}$ have already been pushed to their ``$2
\sigma$" lower bounds so this discrepancy is at greater than the 95\%
confidence level.  The crisis emerges!
\begin{figure}
\protect\centerline{
\epsfxsize=4.5in\epsffile[ 30 200 580 620 ]{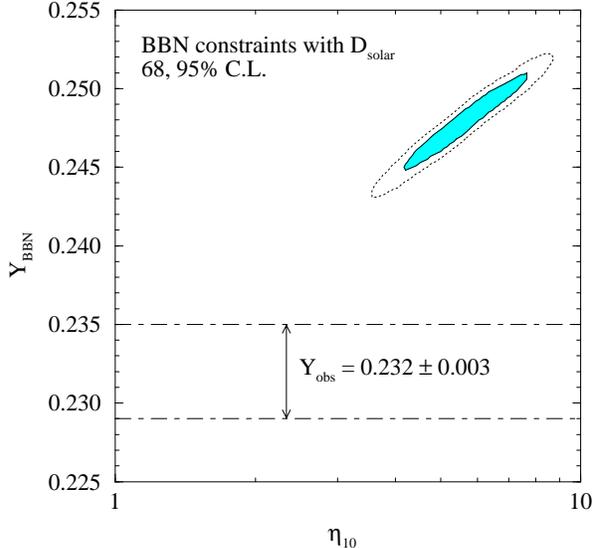}
}
        \caption{The predicted 68\% and 95\% CL contours for the $^4$He primordial mass fraction with $\eta$ constrained by D, $^3$He and $^7$Li.  Also shown is the $\pm 1\sigma$ range for $Y_P$ inferred from the data.}
       
\end{figure}

Indeed, Olive \& Steigman (1995) used all the data (D, $^3$He, $^4$He, $^7$Li)
to infer
$N_\nu = 2.17 \pm 0.27$ which deviates from the standard model value
$(N_\nu = 3)$ by $\sim 3 \sigma$.   This crisis for SBBN is reflected in
Figure 2
where D, $^3$He $(g_3 \ge 1/4)$ and $^7$Li have been used to bound $\eta$,
leading to predictions of $Y_{BBN}$ at the 68\% and 95\% CL.  Unless the
primordial abundance of $^4$He has been systematically underestimated, the
evidence signals a potential crisis for SBBN.

\section{A Statistical Analysis of BBN}

To explore more carefully the consistency of SBBN, my colleagues and I (Hata et
al. 1995;  Thomas et al. 1995) have undertaken the first comprehensive
statistical
analysis of the confrontation between theory and observation.  We have
reexamined the nuclear and weak interactions and their uncertainties and have
performed a Monte Carlo analysis of the BBN predictions.  Indeed, the curves in
Figure 1 reflect the $\pm 1 \sigma$ uncertainties in the predictions (for $N_\nu
=3$) of $Y_{BBN}$, $y_{2P}$, $y_{7P}$ vs. $\eta$.  From our Monte Carlos we derive
$P(A)_{BBN}$, the probability distributions for the predicted BBN abundances
(A).  We have also reexamined the observational data, accounting for the
statistical uncertainties as well as attempting to allow for various systematic
uncertainties which may arise in using the data to infer the distribution,
$P(A)_{OBS}$, of primordial abundances.  These latter uncertainties are not
necessarily modelled by gaussian distributions.  In contrast to previous
approaches which treated each element one at a time, we may use the
information on all light nuclides to form a likelihood function (as a
function of
$\eta$ and $N_\nu$) from $P(A)_{BBN}$ and $P(A)_{OBS}$.  The likelihood
function, maximized with respect to $\eta$ at each $N_\nu$ is shown in Figure
3 (from Hata et al. 1995).  We derive $N_\nu = 2.1 \pm 0.3$, consistent with
Olive \& Steigman (1995).  It is clear that SBBN ($N_\nu = 3$) provides a
poor fit
to the primordial abundances inferred from the data.

\begin{figure}
\protect\centerline{
\epsfxsize=3.5in\epsffile[ 30 80 580 580 ]{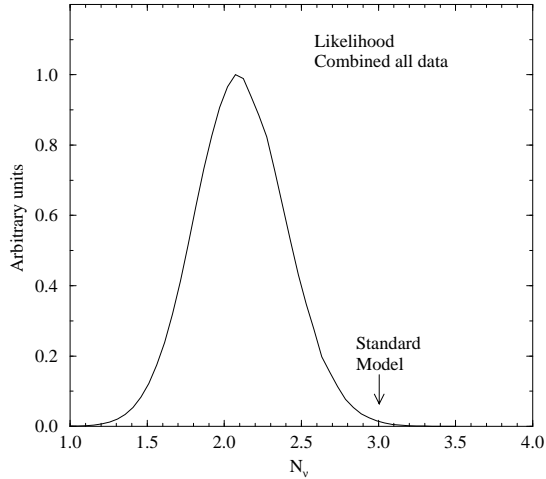}
}
        \caption{The
likelihood function (arbitrary normalization) for the combined fit (D,$^3$He, $^4$He, $^7$Li) of the data and BBN as a function of $N_{\nu}^{BBN}$.  At each value of $N_{\nu}^{BBN}$ the likelihood is maximized for $\eta$.}
       
\end{figure}

As with the 19th century standard model, SBBN is challenged.  As with the
challenges to the 19th century standard model, there are several options for the
resolution of this crisis.

\subsection{Is The Chemical Evolution Extrapolation Wrong?
}

One source of the challenge to SBBN is the relatively high lower bound to $\eta$
imposed by the relatively low primordial abundances of D and $^3$He inferred
from solar system and interstellar observations.  These stringent upper
bounds to
primordial D and $^3$He are suggested by many specific chemical evolution
models (Steigman \& Tosi 1992) as well as the ``generic" model for the evolution
of D and $^3$He (Steigman \& Tosi 1995;  Hata et al. 1996).  In the
latter case,
the crisis worsens with increasing $g_3$ and/or if stellar production of
$^3$He is
allowed for.  The crisis could be ameliorated if $g_3$ is less than the
lower bound
$(g_3 \ge 1/4)$ adopted in the above analyses.  In Figure 4 (from Hata et al.
1995), 95\% CL contours are shown in the $N_\nu$ vs. $\eta$ plane for several
choices of $g_3$.  If the ``effective" $g_3$ (averaged over stars of all
masses and
the evolution history of the ISM) is $\sim 0.1$, consistency of SBBN is
reestablished.
\begin{figure}
\protect\centerline{
\epsfxsize=4.5in\epsffile[ 30 200 580 620 ]{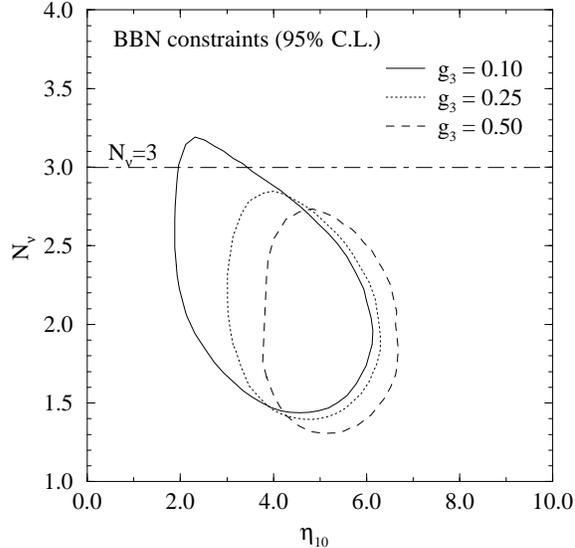}
}
        \caption{The $95\%$ CL contours in the $N_{\nu}^{BBN}$ vs.
$\eta$ plane for several choices of the $^3$He survival fraction g$_3$.}
       
\end{figure}

\subsection{Is The Primordial $^4$He Abundance Larger?}

An alternate source of the challenge to SBBN is the relatively low abundance of
primordial $^4$He inferred from the observations of extragalactic HII
regions.  By
allowing only for statistical uncertainties perhaps we've underestimated
the true
uncertainty in $Y_P$.  A larger value for $Y_P$ could reestablish the
consistency
of $^4$He with D and $^3$He.  Many sources of possible systematic
uncertainty in $Y_P$ have been identified and some have been studied (Davidson
\& Kinman 1985;  Pagel et al. 1992;  Skillman \& Kennicutt 1993;  Skillman
et al.
1994;  Copi, Schramm \& Turner 1995;  Sasselov \& Goldwirth 1995;  Olive \&
Steigman 1995).  In Figure 5 (from Hata et al. 1995) are shown 95\% CL contours
in the $N_\nu$ vs. $\eta$ plane for several choices of $\Delta Y_{sys}$, where
$Y_{BBN} = 0.232 \pm 0.003 + \Delta Y_{sys}$.  If $Y_P$ is shifted up by $\rsim
0.010$, SBBN may be consistent at the 95\% CL.  It should, however, be
emphasized that $\Delta Y_{sys}$ may be negative as well as positive;  a
negative
$\Delta Y_{sys}$ exacerbates the crisis for SBBN.

\subsection{Is There New Physics?}

By employing $N_\nu$ as a second parameter, we have allowed for a class of
modifications of the standard model.  If, in addition to three flavors of light,
left-handed neutrinos ($N_\nu = 3$, SBBN) there are additional light
neutrinos or
other new particles, $N_\nu > 3$ (Steigman, Schramm \& Gunn 1977) and the
crisis worsens.  However, although $\nu_e$ and $\nu_\mu$ are known to be
``light" ($m_\nu < T_{BBN}$), accelerator data on $\nu_\tau$ (ALEPH
Collaboration) permits $m_{\nu \tau} \le 24 MeV$.  As Kawasaki et al. (1994)
have shown, the presence of a massive, unstable tau neutrino with $5 - 10 \lsim
m_{\nu \tau} \le 24 MeV$ and $0.01 \lsim \tau_{\nu \tau} \lsim 1 \, $sec. would
correspond to an ``effective"$N_\nu < 3$.  Perhaps the crisis for SBBN is
teaching us about extensions of the standard model of particle physics.
\begin{figure}
\protect\centerline{
\epsfxsize=4.5in\epsffile[ 30 200 580 620 ]{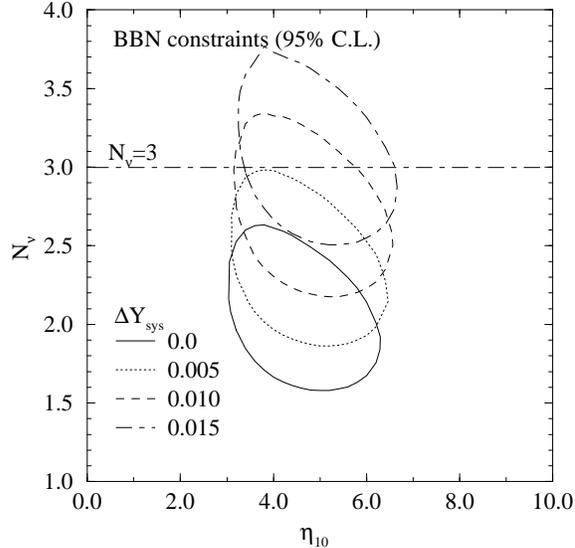}
}
        \caption{The $95\%$ CL contours in the $N_{\nu}^{BBN}$ vs.
$\eta$ plane for several choices of the systematic error ($\Delta{Y_{sys}}$) in the $^4$He abundance inferred from HII region data.}
       
\end{figure}

\section{Summary and Conclusions}

Primordial nucleosynthesis must have occurred during the early, hot, dense
evolution of a Universe described by the hot big bang model.  Therefore, BBN
offers a test of standard cosmology as well as a probe of particle physics.
As with
the standard model of 19th century physics, over many years SBBN has provided
support for the standard model of cosmology.  Indeed, the success of SBBN in
predicting the abundances of the light nuclides with only one adjustable
parameter $\eta$ ($N_\nu =3$) restricted to a narrow range $(3 \lsim \eta_{10}
\lsim 4)$ while the abundances range over some 9 orders of magnitude, is
impressive indeed.

However, as with the 19th century standard model, some clouds have now
emerged on the horizon.  Recent analyses (Copi, Schramm \& Turner 1995;  Olive
\& Steigman 1995;  Hata et al. 1995) point to a crisis unless the data are
in error,
or the extrapolations of the data are in error, or there is new physics.  Some
analogies with the crises which confronted the 19th century standard model may
be instructive.

\noindent (i) Perhaps our extrapolations of the observations of D and $^3$He
from here and now to there and then have been naive.  Chemical evolution
models in which more D is cycled through stars and destroyed (without a
concommitant overproduction of $^3$He) would permit a higher primordial
abundance of D, allowing a lower $\eta$ and $Y_{BBN}$ consistent with
$Y_P$.  Thus, as with the discovery of Uranus, the crisis for SBBN may teach us
something new about galactic evolution.

\noindent (ii) Perhaps our estimates of the primordial abundance of $^4$He are
in error because we have overlooked some large systematic error in the
abundance determinations.  If $Y_P$ is larger than the value inferred form the
observational data, $\eta$ may be as large as inferred from D and $^3$He (see
Figure 2) and still $Y_{BBN}$ and $Y_P$ may be consistent.  Then, as with the
discovery of Pluto, our crisis may have been a false alarm from which,
nonetheless, we learn something new.

\noindent (iii) The most exciting possibility, of course, would be that the
data are
accurate, the systematic errors small and the extrapolations true.  Then,
this crisis
may point us to new physics beyond the standard models of particle physics or
cosmology.  The ``window" on a massive, unstable $\tau$-neutrino is accessible
to current accelerators.

To summarize, then, SBBN ($N_\nu =3; \, g_3 \ge 1/4; \, \Delta Y_{sys} \le
0.005$) provides a poor fit to the primordial abundances of the light nuclides
inferred from current observational data (Hata et al. 1995).  This crisis
is not a
cause for alarm, but an opportunity to learn something new about astronomy,
cosmology, or particle physics.

\acknowledgments

The work described here has been done in collaboration with S. Bludman, N. Hata,
P. Langacker, K. Olive, R. Scherrer, D. Thomas, M. Tosi and T. Walker.  I thank
them for all that I've learned in our collaborations, and for permission to
present
our joint work here.  The research of the author is supported at Ohio State
by the
DOE (DE-AC02-76-ER01545).  I am pleased to thank Steve Holt and his colleagues
and staff for all their support and patience.

\end{document}